\documentclass[a4paper,12pt]{article}

\usepackage[utf8]{inputenc}
\usepackage[english]{babel}
\usepackage{authblk}
\usepackage[dvipdfmx]{graphicx}
\usepackage{mathptmx}
\usepackage[singlespacing]{setspace}
\usepackage[headheight=1in,margin=1in]{geometry}
\usepackage{fancyhdr}
\usepackage{lipsum}

\usepackage{subcaption}
\usepackage{amsmath}
\usepackage{hyperref}

\usepackage{booktabs}
\usepackage{xspace}
\usepackage[dvipdfmx]{color}

\makeatletter
\def\@maketitle{%
  \newpage

  \begin{center}%
  \let \footnote \thanks
    {\LARGE \@title \par}%
  \end{center}%
  \par
  \vskip 0.1em}
\makeatother

\graphicspath{{images/}}

\title{Effect of Monetary Reward on Users' Individual Strategies Using Co-Evolutionary Learning}

\date{}

\begin{document}

\maketitle
\thispagestyle{fancy}

\begin{center}
  Shintaro Ueki$^1$, Fujio Toriumi$^2$ and Toshiharu Sugawara$^1$\\[12pt]
  \hfil
  \begin{minipage}[t]{9cm}
    \centering
    $^1$Department of Computer Science and\\ Communications Engineering\\
    Waseda University\\
    Tokyo 1698555, Japan\\
    s.ueki@isl.cs.waseda.ac.jp, sugawara@waseda.jp
  \end{minipage}
  \hfil
  \begin{minipage}[t]{6.5cm}
    \centering
    $^2$Department of Systems Innovation\\
    The University of Tokyo\\
    Tokyo 1138654, Japan.\\
    tori@sys.t.u-tokyo.ac.jp;
  \end{minipage}\hfil\\[12pt]
\textit{Keywords: Consumer generated media, Game theory, Agent-based simulation,\\ Monetary reward, Co-evolution}
\newline
\end{center}

\section*{Extended Abstract}
\noindent{\bf Background:} {\em Consumer generated media} (CGM), such
as social networking services (SNS) and review sites, are used by many
people; however, these sites rely on the voluntary activity of users
to prosper, garnering the psychological rewards of feeling connected
with other people through comments and reviews received online. To
attract more users, some CGM have introduced monetary rewards (MR) for
publishing activity. However, the effect of MR on the article posting
strategies of users, especially frequency and quality, has not been
fully analyzed. Usui et al.~\cite{Usui2022Scheme} proposed a game
theoretical-model, the {\em SNS-norms game with monetary reward and
  article quality} (SNS-NG/MQ), and investigated the dominant behavior
of users by introducing a few strategies for providing MR (MR
strategies) using the genetic algorithm (GA). They found that although
MR increases the frequency of article posts, the positive or negative
impact on article quality depends on the MR strategy. 
\par

However, the dominant strategies determined by the naive GA are almost
identical for all users. We believe that appropriate strategies for
CGM depend on the standpoints of users, such as normal users or
{\em influencers} who have a large number of followers. However, such
differences were ignored in their study although exploring the
individual behaviors of users with different attitudes is crucial for
CGM to infer the overall behavioral structure of all users.
\par

\vspace{3pt}\par
\noindent{\bf Purpose and Method:} The purpose of this study is to
investigate the effect of MR on individual users by considering the
differences in dominant strategies with respect to user
standpoints. To this end, to determine the individual strategies of
users in SNS-NG/MQ, we applied {\em multiple-world GA}
(MWGA)~\cite{Miura2021MwgaNorms} instead of GA. MWGA generates several
copies of a CGM network with nodes as agents that correspond to users,
and each agent in the multiple worlds selects different behavioral
strategies to interact with its neighboring agents. Then, the agent
with the larger reward is likely to be selected as one parent for the
next generation, spreading the better strategies to agents at the same 
location in many worlds. As a
result, each agent can have its own dominant strategy in CGM. The
reader is referred to \cite{Miura2021MwgaNorms} for details of the
algorithm.
\par

We briefly describe SNS-NG/MQ.
A CGM network represents the connections between agents,
as expressed by graph $G=(V,E)$, where $V=\{1, \dots, N\}$ is the set
of $N$ agents, and $E$ is the set of edges corresponding to friend
relationships. Agent $i\in V$ uses three parameters to describe the
behavioral strategy, (article) posting rate $B_i$, comment rate $L_i$,
and quality $Q_i$ ($0\leq B_i, L_i, Q_i\leq 1$); these values are
decided by MWGA for $i$'s own dominant strategy. Agent $i$
also has a random value, $0\leq M_i\leq 1$, representing the degree of
the user's preference for MR, and is specified at the onset because it
may be innate or acquired but underlying preference.
Then, users are classified into two disjoint sets,
$V_\alpha =\{i \in V \mid M_i < 0.5 \}$ and $V_\beta =\{i \in V \mid
M_i \geq 0.5 \}$, which are the sets of agents preferring
psychological and monetary rewards, respectively.
\par

\begin{figure}
\centering
\includegraphics[width=0.7\hsize]{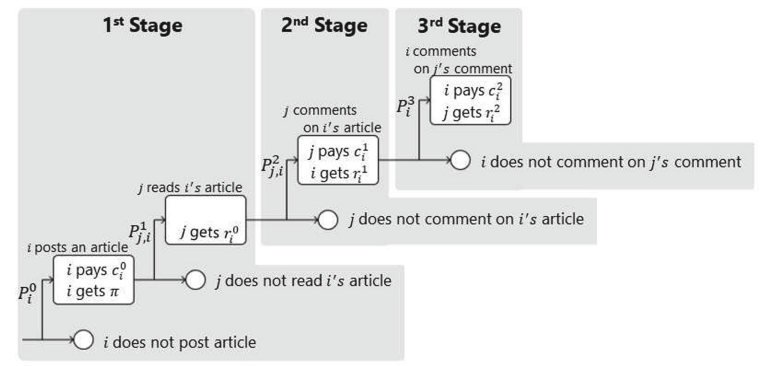}
\caption{Flow of the SNS-norms game with monetary reward and article quality.}
\label{fig:monetaryNorms}
\end{figure}

\begin{table}
\caption{Probabilities used in SNS-NG/MQ (\autoref{fig:monetaryNorms}).}
\label{tab:eqp}
\centering
\begin{tabular}{lll}
  \toprule
  Description &	Parameters & Value \\
  \midrule
  Probability of article post  & $P^0_i$ & $B_i \times Q_{min}/Q_i$ \\
  Probability of article read  & $P^1_{j,i}$  & $Q_i/s_j$ \\
  Probability of comment post  & $P^2_{j,i}$  & $L_j \times Q_i$ \\
  Probability of posting a meta-comment & $P^3_i$  & $L_i \times Q_i$ \\
  \bottomrule
\end{tabular}
\end{table}

\begin{table}
\caption{Calculation of costs, psychological rewards, and utility in SNS-NG/MQ.}
\label{tab:eqcr}
\centering
\begin{tabular}{lll}
  \toprule
  Description & Parameters  & Formula\\
  \midrule
  Cost of article post & $c^0_i$  & $c_{ref}\times Q_i $\\
  Cost of comment      & $c^1_i$  & $c_{ref}\times \delta$ \\
  Cost of meta-comment    & $c^2_i$  & $c^1_i\times \delta$\\
  Psy. reward of article read   & $r^0_i$  & $c^0_i\times \mu$\\
  Psy. reward of receiving a comment  & $r^1_i$  & $c^1_i\times \mu$ \\
  Psy. reward of receiving a meta-comment & $r^2_i$  & $c^2_i\times \mu$\\
  Utility & $u_i$   & $(1-M_i) \times R_i + M_i \times K_i - C_i$\\
 \bottomrule
\end{tabular}
\end{table}

The flow of the game on this graph is illustrated in
\autoref{fig:monetaryNorms}, in which states transit depending on
probabilities, as listed in \autoref{tab:eqp}. The effect of $Q_i$ is
that the articles with high $Q_i$ requires more cost but increases the
probability that the posted article will be read and hence, likely to
receive comments.
A game round is defined as the round in which all agents have a chance
for an article post. Note that $s_j$ is the article number that $i$'s
neighboring agent $j$ can read in a game round. A generation of MWGA
is four game rounds. A number of MR
strategies can be considered for MR $\pi>0$; however, due to the page
limitation, we assume that agent $i$ can receive $\pi$ when $i$
posts an article. Other parameters describing costs and rewards are
identical to those in \cite{Usui2022Scheme}, which are also listed in
\autoref{tab:eqcr}.
\par

\begin{table}
\caption{Parameter values in experiments.}
\label{tab:param}
\centering
\begin{tabular}{lll}
 \toprule
 Description         & Parameters  & Value  \\
 \midrule
 Number of agents        & $N=|V|$  & 400  \\
 Number of agents preferring psychological reward   & $|V_\alpha|$  & 200  \\
 Number of agents preferring monetary reward   & $|V_\beta|$  & 200  \\
 Reference value for cost and psychological reward   & $c_{ref}$  & 1.0  \\
 Raito of cost to psychological reward     & $\mu$  & 8.0  \\
 Cost ratio between game stages     & $\delta$  & 0.5  \\
 Monetary reward        & $\pi$   & 1.0  \\
 Number of worlds in MWGA        & $W$   & 10  \\
 Number of generations       & $g$   & 1000  \\
 Probability of mutation       & $m$   & 0.01  \\ 
 \bottomrule
\end{tabular}
\end{table}

\begin{figure}
\centering
\begin{minipage}[t]{0.45\hsize}
 \centering
 \includegraphics[width=8cm]{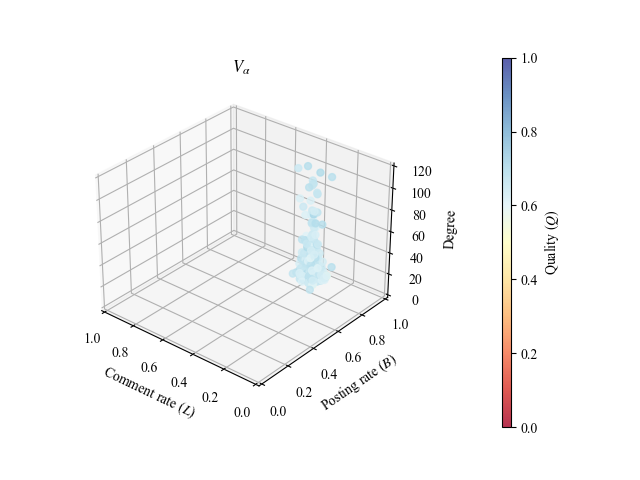}
 \subcaption{Agents preferring psychological reward $V_\alpha$}
 \label{fig:Va1GA}
\end{minipage}
\begin{minipage}[t]{0.45\hsize}
 \centering
 \includegraphics[width=8cm]{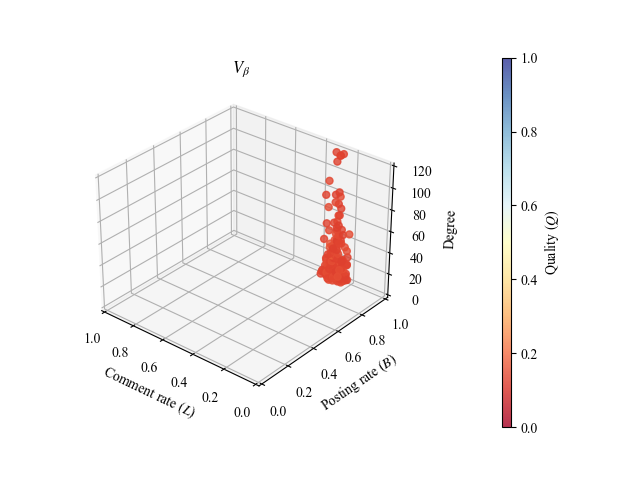}
 \subcaption{Agents preferring monetary reward $V_\beta$}
 \label{fig:Vb1GA}
\end{minipage}
\caption{3D scatter plots of strategy parameters and degree in GA.}
\label{fig:V1GA}
\end{figure}

\begin{figure}
\centering
\begin{minipage}[t]{0.45\hsize}
 \centering
 \includegraphics[width=8cm]{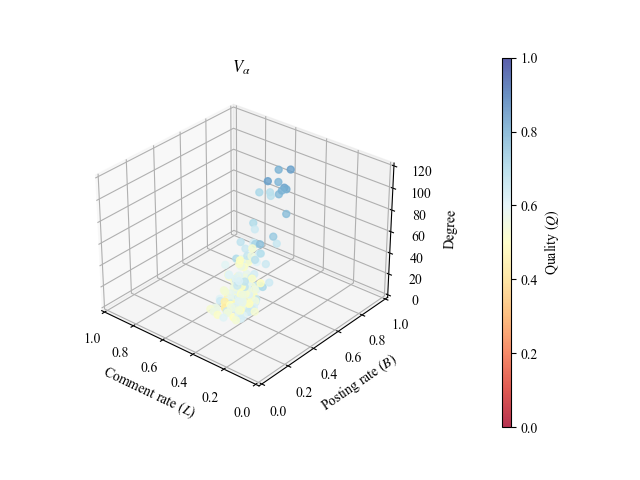}
 \subcaption{Agents preferring psychological reward $V_\alpha$}
 \label{fig:Va1}
\end{minipage}
\begin{minipage}[t]{0.45\hsize}
 \centering
 \includegraphics[width=8cm]{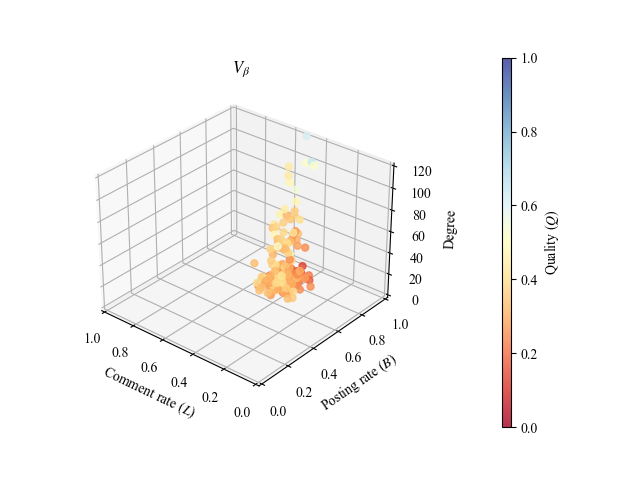}
 \subcaption{Agents preferring monetary reward $V_\beta$}
 \label{fig:Vb1}
\end{minipage}
\caption{3D scatter plots of strategy parameters and degree in MWGA.}
\label{fig:V1}
\end{figure}

%\vspace{3pt}\par
\noindent{\bf Experimental Results and Discussion:}
The experimental setting is the same as in \cite{Usui2022Scheme} and
listed in \autoref{tab:param}, excluding that graph $G$ is generated
using the {\em connecting nearest neighbor model}~\cite{CNN} with the
transition probability from a potential edge to a real edge set to
$u=0.9$, and $N=400$. This model generates a well-known scale-free,
small-world, undirected network with a high cluster coefficient. In
\autoref{fig:V1GA} and \autoref{fig:V1}, the values of $B_i$, $L_i$,
and agent's degree, are plotted for GA~\cite{Usui2022Scheme} and our
method with MWGA, respectively. The heat legend indicates the quality
value $Q_i$.
\par

The results of the previous study (\autoref{fig:V1GA}) suggest
both acceptable and unacceptable behaviors of agents. First, 
agents, especially those in $V_\beta$, attempt to post numerous but
low-quality articles. This behavior seems reasonable. Such articles
receive few comments, and hence low psychological rewards, but the
agents could gain more MRs from posting numerous low-quality
articles. Another observation is that the strategies of all agents
were almost uniform, i.e., they have similar values of $B_i$ $L_i$,
and $Q_i$, regardless of their degrees. This result is
somewhat counterintuitive because in actual CGMs, the strategies of
influencers differ from those of normal users.
\par

Meanwhile, \autoref{fig:V1} indicates that agents have diverse
strategies depending on their standpoints in the networks.
The most remarkable difference is that the influencers, which have
high degrees, behave such that they write as many high
quality articles as possible at a high cost. We believe that this
result is consistent with the actual behaviors of users in CGM.
In particular, they indicate a situation where both psychological and 
monetary rewards are effectively used. In addition, the number of
articles submitted by all agents increases while the quality of
articles decreases. This depends on the value of the monetary reward
$\pi$, and increasing $\pi$ makes the situation worse; low-quality
articles increased.
\vspace{3pt}\par

\noindent{\bf Conclusion:}
Experiments were conducted on SNS-NG/MQ~\cite{Usui2022Scheme} using
MWGA to obtain more diverse realistic dominant strategies depending on
user standpoints in the CGM network. We introduced a number
of MR strategies and conducted the experiments other than those
discussed here. These results will provide insight to CGM
platformers regarding the MR strategies that can make CGMs thrive or
fail. We plan on investigating different models of CGM with other
reward structures.

\bibliographystyle{unsrt}
\bibliography{ref.bib}

\end{document}